# Controlling Synthetic Spin-Orbit Coupling in a Silicon Quantum Dot with Magnetic Field

Xin Zhang,[1,2,#] Yuan Zhou[1,2,#] Rui-Zi Hu,[1,2] Rong-Long Ma[1,2] Ming Ni,[1,2] Ke Wang,[1,2] Gang Luo,[1,2] Gang Cao,[1,2] Gui-Lei Wang,[3] Peihao Huang,[4,5] Xuedong Hu,[6] Hong-Wen Jiang,[7] Hai-Ou Li,[1,2,*] Guang-Can Guo,[1,2] and Guo-Ping Guo[1,2,8*]

[1] *CAS Key Laboratory of Quantum Information, University of Science and Technology of China, Hefei, Anhui 230026, China*

[2] *CAS Center for Excellence and Synergetic Innovation Center in Quantum Information and Quantum Physics, University of Science and Technology of China, Hefei, Anhui 230026, China*

[3] *Key Laboratory of Microelectronics Devices & Integrated Technology, Institute of Microelectronics, Chinese Academy of Sciences, Beijing 100029, China*

[4] *Shenzhen Institute for Quantum Science and Engineering, Southern University of Science and Technology, Shenzhen 518055, China*

[5] *Guangdong Provincial Key Laboratory of Quantum Science and Engineering, Southern University of Science and Technology, Shenzhen, 518055, China*

[6] *Department of Physics, University at Buffalo, SUNY, Buffalo, New York 14260, USA*

[7] *Department of Physics and Astronomy, University of California, Los Angeles, California 90095, USA*

[8] *Origin Quantum Computing Company Limited, Hefei, Anhui 230026, China*

[#] These authors contributed equally to this work.

[*] Corresponding author. Emails: haiouli@ustc.edu.cn (H.-O. L.); gpguo@ustc.edu.cn (G.-P.G.).

## Abstract

Tunable synthetic spin-orbit coupling (s-SOC) is one of the key challenges in various quantum systems, such as ultracold atomic gases, topological superconductors, and semiconductor quantum dots. Here we experimentally demonstrate controlling the s-SOC by investigating the anisotropy of spin-valley resonance in a silicon quantum dot. As we rotate the applied magnetic field in-plane, we find a striking nonsinusoidal behavior of resonance amplitude that distinguishes s-SOC from the intrinsic spin-orbit coupling (i-SOC), and associate this behavior with the previously overlooked in-plane transverse magnetic field gradient. Moreover, by theoretically analyzing the experimentally measured s-SOC field, we predict the quality factor of the spin qubit could be optimized if the orientation of the in-plane magnetic field is rotated away from the traditional working point.

# I. INTRODUCTION

Electron spins in semiconductor quantum dots (QDs) are considered one of the most promising qubit designs for scalable quantum information processing [1-3]. By applying an alternating magnetic field, the electronic spin can be coherently controlled through electron spin resonance (ESR) [4]. Alternatively, such control can be implemented electrically via intrinsic or synthetic spin-orbit coupling (SOC), which is termed as electric-dipole spin resonance (EDSR) [5,6]. In combination with the long spin coherence time in natural silicon, which is further improved by zero-spin-isotope purification, the synthetic spin-orbit coupling (s-SOC) has enabled high-fidelity single-, two-, and multi-qubit operations, as well as strong spin-photon coupling and long-range qubit interactions in Si QDs [7-18].

However, with time inversion asymmetry [19,20], s-SOC also exposes a spin qubit to electric noise and gives rise to fast spin relaxation [21,22] and pure dephasing [7,8,23-25]. Different from the intrinsic spin-orbit coupling (i-SOC) that comes from the underlying atoms and asymmetries in the material or structure, s-SOC in a quantum dot is introduced by a magnetic field gradient from an integrated micromagnet. Concerning the spin quantization axis, this field gradient can be separated into two parts: the transverse component that mediates fast electrical control of spins, and the longitudinal component that adds multi-qubit addressability. In combination with charge noise, the longitudinal field gradient can also cause fast spin dephasing, thus brings uncertainty to the reproducibility and homogeneity of the promised control fidelities [8,14,25]. Therefore, for s-SOC to enable scalable high-fidelity spin qubits in semiconductor QDs, it is crucial to better understand, characterize, and control magnetic field gradients of a micromagnet.

Anisotropy spectroscopy has long been an effective means to probe the physical mechanism of SOC in semiconductor systems [26-34]. Predictably, this method can also be used to investigate s-SOC [31]. In the meantime, transport measurement of ESR or EDSR reveals various physical parameters, such as Larmor and Rabi frequencies, and even spin dephasing times [27,35-38]. Hence, an anisotropy study of transport measured ESR or EDSR should be an effective method to probe the properties of s-SOC. In silicon QDs, there exist valley states that originate from the six-fold degenerate conduction band minimum. The spin and valley degrees of freedom are mixed by spin-orbit coupling [39], whether i-SOC or s-SOC, so that an oscillating electric field can induce simultaneous flip of spin and valley states. This so-called spin-valley resonance [38,40] is different from a normal EDSR that induces transition between Zeeman-split states and offers a conveniently tunable energy gap between spin-valley states at higher magnetic fields for resonance spectroscopy.

Here we report the detection of spin-valley resonance based on the transport measurement of the Pauli spin blockade (PSB) in a natural Si metal-oxide-semiconductor (MOS) double quantum dot (DQD) [1,2]. By controlling the external magnetic field direction in-plane, we find a cosinusoidal modulation of the resonance position with a 180° period and an $8.7 \pm 1.0°$ phase shift. Moreover, a detailed measurement of the resonance peak unveils a strikingly nonsinusoidal modulation of the resonance peak amplitude, which suggests a non-negligible contribution of the in-plane transverse magnetic field gradient of the micromagnet that has long been overlooked in previous studies [9,12,41]. Supported by both the experimental and numerical results, we propose that the s-SOC in semiconductor QDs can be magnetically tuned by rotating the in-plane magnetic field direction, leading to a simultaneous improvement of control rates, dephasing times, and the addressability for spin qubits driven by s-SOC.

## II. RESULTS AND DISCUSSION

### A. Experimental setup

The Si MOS DQD device [34] we study is shown in Fig. 1(a), which is located in a dilution refrigerator with a base temperature ~ 20 mK. Gates C1 and C2 create a channel for electrons to flow between reservoirs under gates L1 (source) and L2 (drain). By selectively tuning gates G1, G2 and G3, a DQD can be defined under gates G1 and G2. Moreover, a rectangular Ti/Co micromagnet of 10 μm by 0.93 μm in the active region (APPENDIX F), with length along the *y*-axis and width along the *x*-axis, as well as a thicknesses of 10/200 nm, is deposited next to the DQD to generate s-SOC with field components $\boldsymbol{B}_p$ parallel to $\boldsymbol{B}_{\text{ext}}$, $\boldsymbol{B}_t$ perpendicular to $\boldsymbol{B}_{\text{ext}}$ and in the *x-y* plane, and $\boldsymbol{B}_z$ perpendicular to both $\boldsymbol{B}_{\text{ext}}$ and the *x-y* plane. Similar to other metal gates, the voltages and the microwave (MW) can also be applied to the micromagnet.

### B. Pauli spin blockade

Our measurement of spin-valley resonance is enabled by the PSB [1] in our DQD. A qualitative sketch of PSB is depicted in the inset of Fig. 1(b) with nominally two electrons. Using S and T to refer to the singlet and the triplet states, respectively, and (1, 1) and (0, 2) to refer to different charge configurations, PSB allows the transition from S(1, 1) to S(0, 2), but not from T(1, 1) to S(0, 2) while interdot detuning $\varepsilon$ is not large enough to make T(0, 2) accessible. The signature of PSB is thus an asymmetric current suppression under bias. As illustrated in Fig. 1(b) in our case, when we measure the current flowing from drain to source, it just corresponds to the electron transiting from (m, n) to (m-1, n+1), where we use m and n to denote the uncertain total electron number in the DQD (see APPENDIX A for the stability diagram with charge sensing), and we find that the leakage current is suppressed in the trapezoidal blockade region inside the two triangles. This process can be intuitively understood using the PSB from (1, 1) to (0, 2) by assuming only valence electron configurations take place in the transport. Also, when we measure the current while varying the energy detuning $\varepsilon$ between (m, n) and (m-1, n+1) and the magnetic field strength, as shown in Fig. 1(c), we can observe the blockade region clearly and obtain a corresponding energy gap of $E_{\text{ST}} \sim 1$ meV. At low field ($B_{\text{ext}} \leq 100$ mT), PSB is partially lifted due to spin-flip cotunneling [42]; while at $B_{\text{ext}}$ in the range of 844 to 896 mT, PSB is lifted due to spin-valley mixing in one of the QDs [38] (see discussion below).

## C. Detection of EDSR.

By setting $V_{G1}$ and $V_{G2}$ within the PSB region and applying continuous microwave (CW) to the micromagnet [13], we measure the transport current $|I_{SD}|$ as a function of both the external magnetic field strength $\boldsymbol{B}_{ext}$ and the microwave frequency $f$. When the spin-valley states are tuned into resonance with the microwave excitation, PSB could be lifted and result in an increased current. In Fig. 2(a), three lines of increased current are visible. The central vertical line corresponds to line V in Fig. 1(c), while two oblique lines A and B on both sides can be understood by the same spin-valley mixing mechanism [38]. As shown in the energy level spectrum of Fig. 2(a), with an increasing magnetic field, two lowest valley states with a valley splitting $E_{VS}$ are split by Zeeman energy $E_Z$, resulting in four spin-valley product states, namely $|1\rangle = |v_-, \downarrow\rangle$, $|2\rangle = |v_-, \uparrow\rangle$, $|3\rangle = |v_+, \downarrow\rangle$ and $|4\rangle = |v_+, \uparrow\rangle$. In the presence of SOC in general, and s-SOC in particular, states $|2\rangle$ and $|3\rangle$ (or $|1\rangle$ and $|4\rangle$) would mix with each other, resulting in two hybridized spin-valley states (APPENDIX D) with an s-SOC strength $\Delta_{SSO}$ indicating the energy gap at the anticrossing of the two states (energy levels of states $|1\rangle$ and $|4\rangle$ never cross, thus their mixing is always relatively small). Therefore, with the oscillating electric field moving the electrons back and forth, the spin state of an electron could be flipped along with its valley state, lifting PSB and thus leading to the observed resonance lines A and B in Fig. 2(a) [38,40].

## D. Anisotropy spectroscopy of spin-valley resonance.

We now focus on the anisotropy of spin-valley resonance. As shown in Fig. 2(b) and (c), by rotating the in-plane magnetic field $\boldsymbol{B}_{ext}$ with an angle $\phi$ with respect to the $x$-axis and keeping the microwave frequency constant at 10.09 GHz, we scan the strength of the external magnetic field for resonances A and B and find they are modulated by the field orientation. Without loss of generality, we take resonance B as an example to perform a detailed study of the anisotropic resonance position and resonance amplitude $I_p$, as shown in Fig. 3(a) and (b), with both quantities extracted by fitting the resonance peak with a Gaussian function [1] [inset of Fig. 3(a)].

Fig. 3(a) shows a cosinusoidal modulation of resonance position with a 180° period and an $8.7 \pm 1.0°$ phase shift. To make a comparison, we calculate the stray magnetic fields along different directions generated from the micromagnet. In particular, $\boldsymbol{B}_p$ (the solid dark blue curve), which is parallel to $\boldsymbol{B}_{ext}$, shows nearly out-of-phase modulation compared to the resonance peak positions. This negative correlation can be understood by the fact that the direction of the total magnetic field is nearly along $\boldsymbol{B}_{ext}$ and thus $\boldsymbol{B}_p$ contributes most through $hf \sim \gamma(\boldsymbol{B}_{ext} + \boldsymbol{B}_p)$, where $h$ is the Planck constant, $f$ is the fixed microwave frequency we applied, and $\gamma$ is the gyromagnetic ratio. Moreover, such a relationship between $\boldsymbol{B}_{ext}$ and $\boldsymbol{B}_p$ suggests s-SOC dominates the anisotropy over i-SOC in our device. Our numerical calculation also indicates that the small phase shift of the cosinusoidal curve is caused by the deviation of the electron position from the centerline along the length of the rectangular micromagnet.

In contrast, Fig. 3(b) shows a nonsinusoidal modulation of resonance amplitude $I_p$, though with the same period and similar modulation phase as the resonance position. This behavior is radically different from the sinusoidal anisotropy due to i-SOC shown in previous work [38], and likely originates from s-SOC. To a first approximation, $I_p$ is proportional to the square of Rabi oscillation rate $\omega_R$ [27,35,36], and by deriving the equation for $\omega_R$ in the limit of $|E_{\text{VS}} - E_z| \gg |\Delta_{\text{SSO}}|$ (APPENDIX D), we get:

$$I_p = C b_{tr}^2 \qquad (1)$$

where $b_{tr}$ is the transverse magnetic field gradient along the electron displacement direction and the origin of s-SOC strength $\Delta_{\text{SSO}}$, while $C$ is a constant scaling factor. The total magnetic field direction $\boldsymbol{B}_{\text{tot}} = \boldsymbol{B}_{\text{ext}} + \boldsymbol{B}_p + \boldsymbol{B}_t + \boldsymbol{B}_z$ defines the exact spin quantization axis, and the electron displacement direction is along the $y$-axis. Thus the total transverse magnetic field gradient should be $b_{tr} = d\boldsymbol{B}_{\text{tr}}^{\text{tot}}/dy$. We have numerically calculated $I_p = C\,(d\boldsymbol{B}_{\text{tr}}^{\text{tot}}/dy)^2$, and it reproduces the basic features of the experimental results quite well [see the navy curve in Fig. 3(b)].

The calculated $I_p$ curve may be counterintuitive at the first sight. With an intuitive picture of the magnetic induction lines from the rectangular micromagnet, one would normally expect that the maximal $I_p$ is along the length ($\phi = 90°$ or $270°$, $y$-axis) of the micromagnet and the minimal $I_p$ along the width ($\phi = 0°$ or $180°$, $x$-axis). However, as shown in Fig. 3(b), though the angle of minimal $I_p$ is as expected, the angles of maximal $I_p$ deviate from the $y$-axis significantly, and $I_p$ has two peak values in a single period. To explain this phenomenon, we calculate the resonance amplitudes induced by the in-plane ($d\boldsymbol{B}_{\text{tr}}^{\text{in}}/dy$) and out-of-plane ($d\boldsymbol{B}_{\text{tr}}^{\text{out}}/dy$) transverse magnetic field gradients separately (see Fig. 4(a) for different magnetic field gradients). As shown in Fig. 3(b), $d\boldsymbol{B}_{\text{tr}}^{\text{out}}/dy$, with the maximum value near the $y$-axis and a cosinusoidal curve of $180°$ period, is in good agreement with the intuitive expectation. However, $d\boldsymbol{B}_{\text{tr}}^{\text{in}}/dy$, though is usually neglected at the traditional working angle [9,12,41] (along the length of the micromagnet), contributes to the total $I_p$ nonnegligibly for certain angles. The nonsinusoidal behavior of the resonance amplitude is a direct result of the competition of the out-of-plane and in-plane transverse magnetic field gradient contributions to the s-SOC.

### E. Optimization of spin control.

In principle, in a resonance experiment dephasing times could be extracted directly from the peak width [37]. However, in our experiment, the microwave power is not low enough to avoid power broadening, and we cannot directly estimate the dephasing times. To circumvent this problem, we calculate the anisotropy of the longitudinal magnetic field gradient $dB_{\text{long}}/dy$ and $dB_{\text{long}}/dx$, which, together with charge noise, should be the most important source for dephasing in our device (APPENDIX E) [8,25]. Interestingly, as shown in Fig. 4(b), we find that when $d\boldsymbol{B}_{\text{tr}}^{\text{tot}}/dy$ approaches its maximum away from the $y$-axis, $dB_{\text{long}}/dy$ decreases to nearly half of its peak value.

In other words, a finite angle away from the $y$-axis for the external field may result in a simultaneous optimization of the dephasing time and the operation rate of the spin-valley qubit. Considering that the transverse and longitudinal gradients are responsible for Rabi oscillation and dephasing respectively and assuming that the charge noise is isotropic, we define a quality factor $Q = (d\boldsymbol{B}_{\text{tr}}^{\text{tot}}/dy)/\sqrt{(d\boldsymbol{B}_{\text{long}}/dy)^2 + (d\boldsymbol{B}_{\text{long}}/dx)^2}$. From this ratio we find that the best angle with the highest control fidelity is around 34° or 161° for our device. Along with these directions, the longitudinal gradient $d\boldsymbol{B}_{\text{long}}/dy$ is severely suppressed while the transverse gradient $d\boldsymbol{B}_{\text{tr}}^{\text{tot}}/dy$ is kept relatively high so that the qubit quality factor is optimized. Moreover, the calculated $d\boldsymbol{B}_{\text{long}}/dx$, which could also be used for spin addressability in our device, shows that it is also enhanced at the angle with the highest $Q$-factor. In short, by aligning the external field away from the electric field direction, we can simultaneously maximize the speed of EDSR for a qubit, minimize its dephasing, while maintaining its addressability.

Compared with i-SOC, which could be strongly influenced by microscopic features of the interface that are difficult to control [33,34], s-SOC is mainly dependent on the micromagnet design whose properties can be reliably predicted by numerical calculations (APPENDIX F) [31]. Therefore, to optimize spin control, most studies focus on how to improve the micromagnet design [24,43,44]. Here, our results suggest that the external magnetic field orientation is another approach to optimize the control fidelity for a spin qubit. Furthermore, while the design of a micromagnet is fixed as soon as it is deposited, external field orientation is tunable in situ. The overall performance of a qubit array can be optimized by rotating the external magnetic field during calibration, making the design and control of a large array of qubits more flexible and effective [45-47].

## III. CONCLUSION

In summary, we have investigated the anisotropy of s-SOC by measuring the spin-valley resonance under a rotating magnetic field. The distinctive nonsinusoidal anisotropy of resonance amplitudes compared to i-SOC shows the significance of the in-plane transverse magnetic field gradients in determining the anisotropy of s-SOC. The calculation of the longitudinal magnetic field gradients also suggests a way to simultaneously optimize the operation rate, the dephasing time, and the addressability of spin qubits by controlling the magnetic field direction. Moreover, our spectroscopy method that employs anisotropic spin resonance to probe s-SOC, with the advantage that can reflect different quantum properties through a single resonance peak, is generally applicable to other quantum systems and semiconductor nanostructures with i-SOC and/or s-SOC, such as one- and two-dimensional material [48,49], topological superconductors [50], etc.

## Acknowledgments:

This work was supported by the National Key Research and Development Program of China (Grant No.2016YFA0301700), the National Natural Science Foundation of China (Grants No. 12074368, 12034018, 11625419, 61922074, 62004185, and 11904157), the Strategic Priority Research Program of the CAS (Grant No. XDB24030601), the Anhui initiative in Quantum Information Technologies (Grants No. AHY080000), the Fundamental Research Fund for the Central Universities (Grants No. WK2030000027), and the Guangdong Provincial Key Laboratory (Grant No. 2019B121203002). H.-W. J. and X. H. acknowledge financial support by U.S. ARO through Grant No. W911NF1410346 and No. W911NF1710257, respectively. This work was partially carried out at the USTC Center for Micro and Nanoscale Research and Fabrication.

## APPENDIX A: CHARGE SENSING

Fig. 5(a) shows the typical bias triangles we measured in the transport regime, of which the triangle in the white dashed rectangle area is the one we measured in the main text. To determine the exact electron number in this area, we use a single-electron transistor (SET) to measure the charge stability diagram under similar conditions. As shown in Fig. 5(b), the irregular resonance tunneling lines hinder an accurate estimate of the electron number under gates G1 and G2. However, we are confident that our experiment was done in the few-electron regime. While our DQD may not have been in the two-electron regime, it experiences the same asymmetric current suppression that is the signature of the two-electron Pauli Spin Blockade (PSB), which can be lifted by spin-flip transitions and has been used for spin measurement [51,52]. Thus we could measure spin-valley resonance, and explain our observation of resonances as the lift of PSB. Moreover, although the valley states in silicon may also complicate the scenario of PSB, the spin-valley blockade could be used similarly to PSB to explain the blockade phenomenon for spin and spin-valley resonance experiments [37,38]. For convenience, we use the same terminologies of the simple (1, 1)- (0, 2) PSB case in the main text.

## APPENDIX B: PSB MEASUREMENT DETAILS

The dc gate voltages are supplied by a 16-channel voltage source, and the continuous microwave is generated by a vector source generator (Keysight E8267D) with -5 dBm power at the output. The microwave transmission line consists of a 13 dB attenuator at room temperature and a 10 dB attenuator at base temperature. The current through source and drain is amplified with a room temperature low-noise current preamplifier (Stanford Research Systems SR570) and measured by a multimeter (Keysight 34410A).

The lever arm of a gate can be extracted from bias triangles. As shown in Fig. 6, since the bias voltage is set at $V_{\text{SD}} = -2$ mV, the lever arm of each gate can be extracted as [53]:

$$\alpha_1 = \frac{e|V_{\text{SD}}|}{\delta V_{\text{G1}}} = 0.333 \text{ eV/V}$$

$$\alpha_2 = \frac{e|V_{\text{SD}}|}{\delta V_{\text{G2}}} = 0.449 \text{ eV/V}$$

Using these lever arms, we obtain $E_{\text{ST}} = 1.056$ meV, and the tunability ~5.96 ueV/meV of valley splitting as a function of $\varepsilon$ in the main text.

## APPENDIX C: EDSR MEASUREMENT DETAILS

In Fig. 2(a) in the main text, there are blank regions with data cleared for clarity. Here we show it completely and also include the intravelly spin resonance line (line I) in Fig. 7(a). In Fig. 7(a), the resonance lines are nearly invisible due to the high leakage current at some frequencies, especially the regions we cleared in Fig. 2(a), thus we reduced the maximum current of the colorbar and reproduced it in Fig. 7(b) to show the data more clearly. The high leakage current in those microwave frequencies should be caused by the excessive microwave power applied, which is due to the uneven microwave transmission to the device for different frequencies. The origin of this inhomogeneity may be the frequency-dependent power attenuation in the a.c. lines and bonding wires we used.

For the intravelly spin resonance, as shown in Fig. 7(b) and (c), we have also measured its anisotropy by scanning the magnetic field strength while keeping the microwave frequency at 10.09 GHz. It can be seen that the intravalley spin resonance I is also cosinusoidally modulated with a phase similar to the spin-valley resonance line A and B, although the magnitude is even smaller. This can be understood that the same s-SOC should also dominate i-SOC for intravalley spin resonance anisotropy and the incomplete magnetization at low fields reduces the anisotropy magnitude. Given that the background leakage current at low magnetic fields is strong due to spin-flip cotunneling (see Fig. 1(c)) and the incomplete magnetization is hard to simulate, we did not explore it in detail to investigate s-SOC but used resonance line B as mentioned in the main text.

In the Gauss fit of spin-valley resonance peaks in the main text, we also extracted the peak baseline (background leakage current) and peak width (full width at half maximum, FWHM) to estimate their anisotropy. As shown in Fig. 7(d), the anisotropy of the peak baseline resembles that of the resonance amplitude but with a much smaller variation magnitude, and the peak width is nearly isotropic, which should be caused by power broadening.

Moreover, for the data acquisition in Fig. 2(b)-(c), Fig. 3(a)- (b), and Fig. 7(c), we have collected the data by rotating the external field up to 720 degrees and more (along the same clock direction), and we did not find any clear hysteresis related to the micromagnet. We think it can be explained by the nearly full magnetization of the micromagnet in the magnetic field range we applied.

# APPENDIX D: THEORETICAL MODEL

Here we propose a model to describe spin-valley resonance in a silicon quantum dot [38,39,54]. As shown in Fig. 2(b) in the main text, only states $|2\rangle = |v_-, \uparrow\rangle$ and $|3\rangle = |v_+, \downarrow\rangle$ are involved in spin-valley resonance. For s-SOC, the total Hamiltonian reads:

$$H = \begin{bmatrix} E_- + \frac{1}{2}E_z & \frac{1}{2}\Delta_{SSO} \\ \frac{1}{2}\Delta_{SSO}{}^* & E_+ - \frac{1}{2}E_z \end{bmatrix} \quad (1)$$

Here, $E_{-(+)}$ refers to the eigenenergy of the corresponding valley state, and $\pm\frac{1}{2}E_z$ depicts their energy shift due to Zeeman splitting under the external magnetic field. The nondiagonal term $\Delta_{SSO} = g\mu_B b_{tr} r_{-+}$ is the strength of s-SOC caused by the transverse magnetic field gradient from the micromagnet, with $g$ the electron g-factor, $\mu_B$ the Bohr magneton, $b_{tr}$ the transverse magnetic field gradient along the electron oscillation direction, and $r_{-+}$ the intervalley transition element.

Diagonalizing the Hamiltonian, we obtain eigenenergies

$$E_{\tilde{2}} = \frac{1}{2}E_{VS} - \frac{1}{2}\varepsilon \quad (2)$$

$$E_{\tilde{3}} = \frac{1}{2}E_{VS} + \frac{1}{2}\varepsilon \quad (3)$$

where $\varepsilon = \sqrt{(E_{VS} - E_z)^2 + \Delta_{SSO}{}^2}$, and the eigenstates

$$|\tilde{2}\rangle = \cos\frac{\theta}{2}|2\rangle - \sin\frac{\theta}{2}|3\rangle \quad (4)$$

$$|\tilde{3}\rangle = \sin\frac{\theta}{2}|2\rangle + \cos\frac{\theta}{2}|3\rangle \quad (5)$$

where

$$\sin\frac{\theta}{2} = \sqrt{\frac{1+a}{2}} \quad (6)$$

$$\cos\frac{\theta}{2} = \sqrt{\frac{1-a}{2}} \quad (7)$$

with

$$a = \frac{E_{VS} - E_z}{\sqrt{(E_{VS} - E_z)^2 + \Delta_{SSO}{}^2}} \quad (8)$$

Assuming the ac electric potential takes the form $V(t) = 2eE_{ac}\cos(2\pi ft)r$, where $e$ is the electron charge, $E_{ac}$ the electric field amplitude, $f$ the oscillation rate, and $r$ the position operator, the total Hamiltonian reads:

$$H_{tot} = \frac{1}{2}\begin{bmatrix} -\varepsilon & V(t) \\ V(t) & \varepsilon \end{bmatrix} \quad (9)$$

Considering the rotating wave approximation under $V(t)$, $H_{tot}$ can be wrriten as:

$$H_{rot} = \frac{1}{2}\begin{bmatrix} -\varepsilon + hf & \hbar\omega_R \\ \hbar\omega_R & \varepsilon - hf \end{bmatrix} \quad (10)$$

with the Rabi frequency

$$\omega_R = \frac{eE_{ac}|F_{SV}||r_{--}-r_{++}|}{\hbar} \quad (11)$$

where

$$F_{SV} = \frac{|\Delta_{SSO}|}{2\sqrt{(E_{VS}-E_z)^2+\Delta_{SSO}^2}} \quad (12)$$

Note that Eq. (11) only differs from the result for intravalley spin resonance [38] by replacing $|r_{-+}|$ with $|r_{--} - r_{++}|$. When $|E_{VS} - E_z| \gg |\Delta_{SSO}|$, which is the case of our experiment, we can obtain

$$F_{SV} \approx \frac{|\Delta_{SSO}|}{2|E_{VS}-E_z|} \quad (13)$$

Therefore, the Rabi frequency is proportional to the s-SOC strength and the transverse magnetic field gradient:

$$\omega_R \approx \frac{eE_{ac}\Delta_{SSO}|r_{--}-r_{++}|}{2|E_{VS}-E_z|\hbar} = \frac{eE_{ac}g\mu_B|r_{-+}||r_{--}-r_{++}|}{|E_{VS}-E_z|\hbar}b_{tr} \quad (14)$$

Since $I_p$ is proportional to $\omega_R^2$, we could obtain the relationship $I_p = Cb_{tr}^2$ in the main text. A comparison of $F_{SV}$ based on Eq. (12) and Eq. (13) are shown in Fig. 8. Assuming an s-SOC strength $|\Delta_{SSO}| = g\mu_B b_{tr}|r_{-+}| \sim 90$ neV, where we use the largest simulated magnetic field gradient $b_{tr} = 0.4$ mT/nm and an estimate of the dipole size $|r_{-+}| = 2$ nm [39], and using the experimental value $E_{VS} = 102.66$ μeV, we find the approximate solution is suitable to describe the data in Fig. 3 in the main text.

For the derivation of the relationship between $I_p$ and $\omega_R$, it can be obtained by finding the steady-state solution of the master equation:

$$\frac{d\rho}{dt} = -\frac{i}{\hbar}[H_{rot}, \rho] + L(\rho) \quad (15)$$

where the Lindblad operator can be written as:

$$L(\rho) = \begin{bmatrix} \Gamma_1\rho_{11} & -\Gamma_2\rho_{01} \\ -\Gamma_2\rho_{10} & -\Gamma_1\rho_{11} \end{bmatrix} \quad (16)$$

with $\Gamma_1$ the longitudinal relaxation rate and $\Gamma_2$ the transverse relaxation rate. By solving the rate equations of $\frac{d\rho}{dt} = 0$, we can obtain:

$$\rho_{11} = \frac{1}{2}\frac{\omega_R^2}{\omega_R^2 + \Gamma_1\Gamma_2 + \left(\frac{\Gamma_1}{\Gamma_2}\right)(\varepsilon-hf)^2} \quad (17)$$

Here $\rho_{11}$ represents the density of states with spin flipped by the microwave excitation, and it contributes to the resonance current by $I_p = e\Gamma_i\rho_{11}$, with $\Gamma_i$ referring to the interdot tunneling rate. Since the experiment in this work was performed within the PSB region and under continuous microwave excitation, the strong decoherence induced by tunneling events will cause $\Gamma_1\Gamma_2 \gg \omega_R^2$. Therefore, $\rho_{11}$ and thus $I_p$ is proportional to $\omega_R^2$ when the qubit is on resonance ($\varepsilon - hf = 0$).

## APPENDIX E: EFFECTS OF MAGNETIC FIELD GRADIENTS

The synthetic spin-orbit coupling consists of transverse and longitudinal components [8], with the transverse components mediating spin rotations driven by an electric field (EDSR), and the longitudinal components contributing to dephasing in combination with fluctuating electrical fields (charge noise). The transverse field gradient is defined by $b_{tr} = (\overrightarrow{e_{MW}} \cdot \boldsymbol{\nabla})B_{MM}^{\perp}$, where $\overrightarrow{e_{MW}}$ is the unit vector along the in-plane oscillating electric field, $\nabla$ is the gradient operator, and $\perp$ denotes the direction perpendicular to $\boldsymbol{B}_{tot}$. Similarly, the longitudinal field gradient is defined by $b_{long} = (\overrightarrow{e_{noise}} \cdot \boldsymbol{\nabla})B_{MM}^{\parallel}$, where $\overrightarrow{e_{noise}}$ is the unit vector along the in-plane fluctuating electric field from the noise, and $\parallel$ denotes the field component parallel to $\boldsymbol{B}_{tot}$. In our experiment, the electrons are strongly confined in the $x-y$ plane in the form of a two-dimensional electron gas (2DEG), while the applied continuous microwave pushes the electrons back and forth along the $y$ direction. The transverse and longitudinal field gradients are therefore defined by $d\boldsymbol{B}_{tr}/dy$, $d\boldsymbol{B}_{long}/dx$ and $d\boldsymbol{B}_{long}/dy$ respectively in the main text. Moreover, since the quantum dots line up along the $x$ direction, the longitudinal field gradient $d\boldsymbol{B}_{long}/dx$ also provides addressability of qubits in different quantum dots.

In the main text, we define a quality factor $Q$ for spin qubit control by the ratio of the transverse and the longitudinal magnetic field gradients. A more common definition $Q^{Rabi}$ is the ratio of Rabi frequency and spin dephasing rate [8,20]. To estimate the advantage of rotating the magnetic field direction, we separate the spin dephasing rate into two parts [33]:

$$\frac{1}{T_2} = \frac{1}{T_2^{sSOC}} + \frac{1}{T_2^{other}} \qquad (18)$$

where $\frac{1}{T_2^{sSOC}}$ is due to s-SOC in combination with charge noise, while $\frac{1}{T_2^{other}}$ comes from other noises such as magnetic noise from residual nuclear spins. From Fig. 4(b) in the main text, we know that when $Q$ is optimized, $\frac{1}{T_2^{sSOC}}$ is severely suppressed and Rabi frequency is kept almost unchanged. Therefore, the improvement of $Q^{Rabi}$ concerning the traditional working point can be approximated by the ratio of $(\frac{1}{T_2^{sSOC}} + \frac{1}{T_2^{other}})/\frac{1}{T_2^{other}}$. If we suppose $T_2^{sSOC} \sim 20$ μs and $T_2^{other} \sim 100$ μs according to the previous results [8,25,55], then the improvement of quality factor by rotating the external magnetic field direction would be about 6 times.

## APPENDIX F: SIMULATION DETAILS

We use the Radia package of Mathematica to simulate the stray field of the micromagnet, assuming a uniform saturation magnetization [6] $M = 1.8\,\text{T}$. The geometry of the micromagnet and its positional relationship with the electron used for simulation are estimated based on the scanning electron microscopy (SEM) image of the device in use, which are summarized in Fig. 9. The micromagnet in the active region has a simple bar magnet geometry, with a width of 930 nm and a length of 10 μm, as shown in Fig. 9(a). Note that the vast majority part of the micromagnet that is beyond the active region and extends to the bonding area is not included, which has little effect on the simulation results and the related conclusions in the main text.

We assume that the electron spin on resonance is underneath gate G2 and the depth is estimated to be equal to the total thickness of the Ti layer (10 nm) and the $SiO_2$ layer (10 nm).

As discussed in the main text, we also simulate magnetic field gradients of the micromagnets of other designs, which are summarized in Fig. 10. Inevitably, the gradients and anisotropy become more complicated for a complex micromagnet design. It is thus of great importance to calculate and check the anisotropy of the magnetic field gradient before performing real experiments and optimize it by controlling the magnetic field direction.

**Figure Captions**

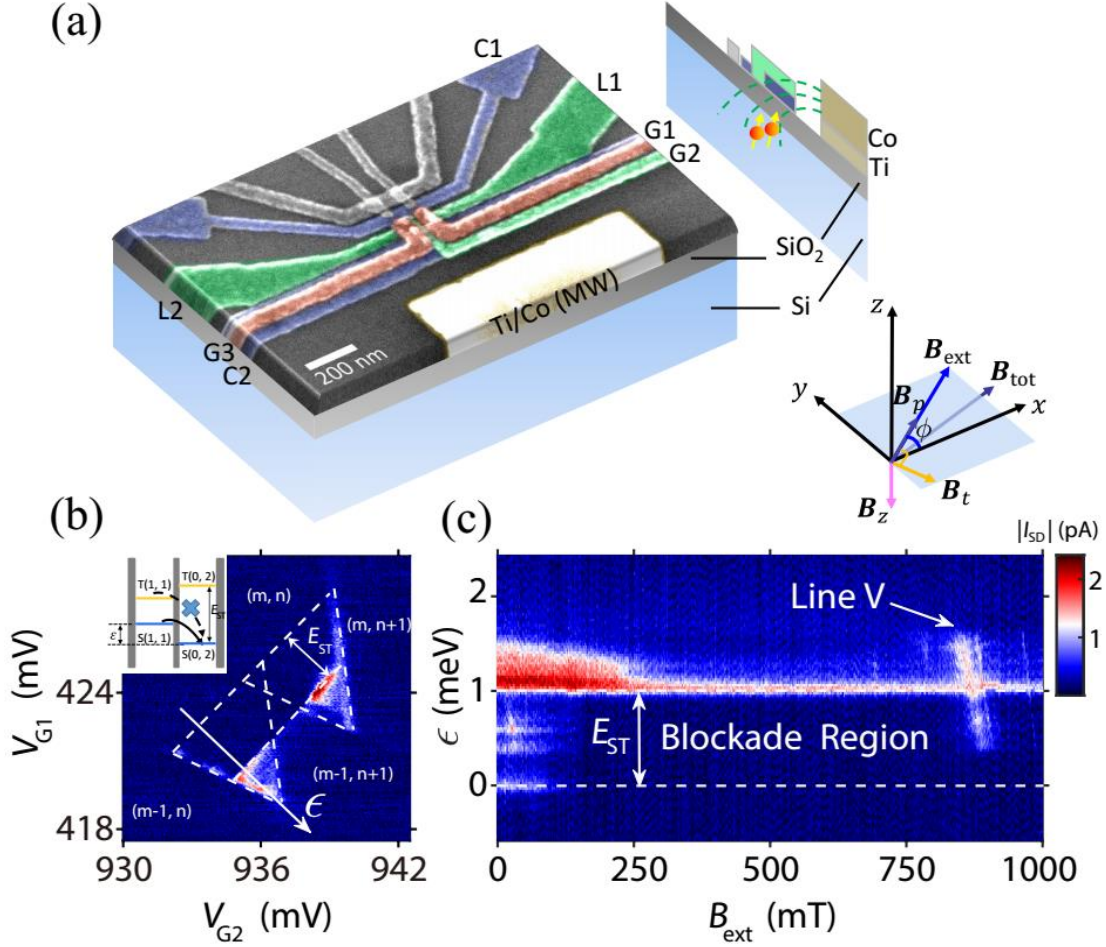

FIG. 1. (a) Schematic of the device layout. The aluminum electrodes and the bar micromagnet used in the experiment are in false colors. Inset: Cartesian coordinate and labels for different magnetic fields with the angle $\phi$ referring to the in-plane orientation of $\boldsymbol{B}_\text{ext}$. (b) Transport current $|I_\text{SD}|$ as a function of $V_\text{G1}$ and $V_\text{G2}$ with a bias voltage $V_\text{SD} = -2$ mV and an external magnetic field $B_\text{ext} = 200$ mT along the $y$-axis (i.e. $\phi = \pi/2$). The PSB results in a current suppression in the bias triangles with a blockade region indicated by an energy gap $E_\text{ST}$ between the two dashed lines. Inset: schematic of the energy levels involved in the PSB, where the delocalized states S(1, 1) and T(1, 1) are only weakly split by exchange interaction and the localized states S(0, 2) and T(0, 2) are split by a much larger energy $E_\text{ST}$ involving an orbital excitation of the QD under gate G2. (c) Transport current $|I_\text{SD}|$ as a function of detuning $\varepsilon$ and external magnetic field $B_\text{ext}$, with the detuning axis highlighted by a white arrow in (b). The blockade region with an energy gap $E_\text{ST}$ between the two dashed lines is also denoted. The leakage current due to spin-valley mixing is labeled by line V. Note line V has a slope ~5.96 ueV/meV of valley splitting with respect to $\varepsilon$ (APPENDIX B), which may be caused by the strong dependence of valley splitting on the electric field under gate G1 or G2.

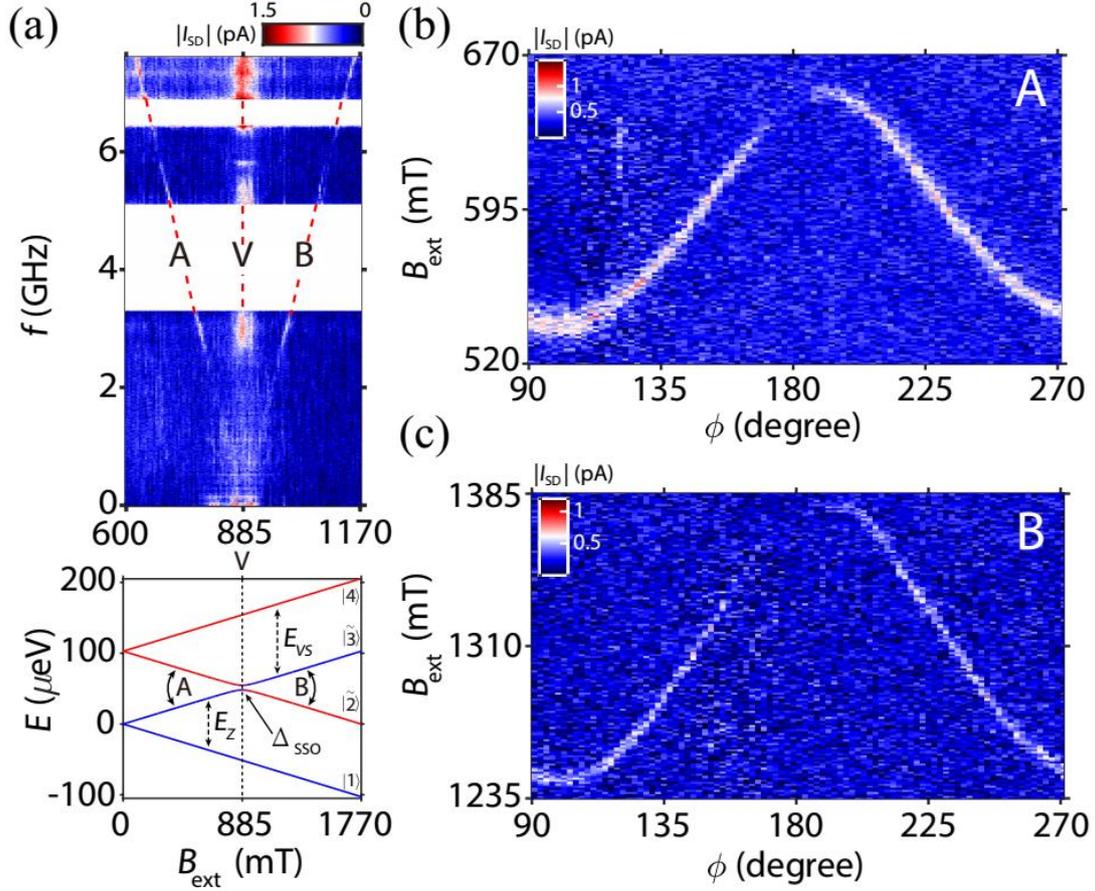

FIG. 2. (a) Transport current $|I_{SD}|$ as a function of the external magnetic field $B_{ext}$ and microwave frequency $f$. Red dashed lines denote the resonant lines where PSB is lifted by the driven spin-flip transition. Data with high leakage current background are cleared for clarity (blank regions) (APPENDIX C). The bottom diagram shows the calculated energy levels for spin-valley mixing. The spin and valley composition of the hybridized states $|\tilde{2}\rangle$ and $|\tilde{3}\rangle$ is indicated by the varied color of the corresponding lines near the anticrossing. Two double-headed arrows mark the corresponding spin-valley transitions A and B. Panels (b) and (c) show the transport current $|I_{SD}|$ as a function of the magnetic field strength $B_{ext}$ and the magnetic field orientation $\phi$ for the resonance A and B, respectively. Notice the anisotropy magnitude of line A (112 mT) is a little smaller than line B (134 mT), which may be attributed to the incomplete magnetization of the micromagnet under lower applied fields.

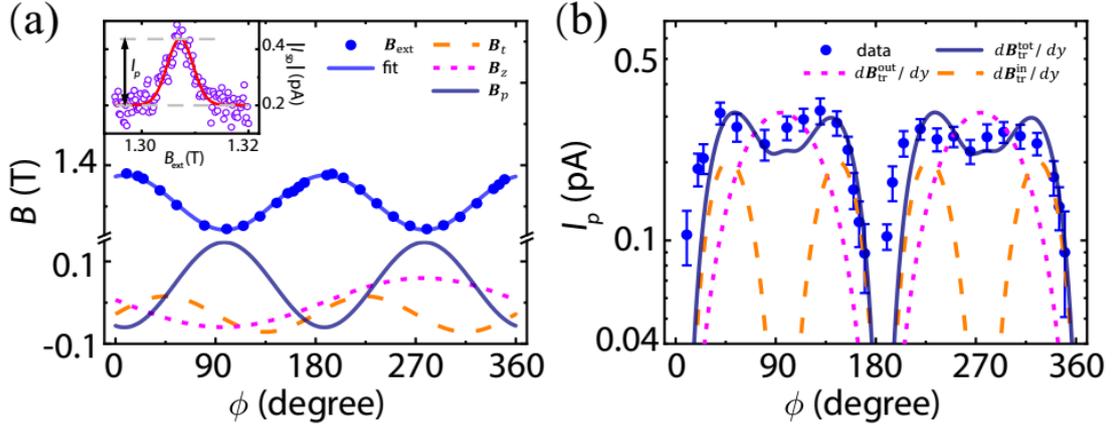

FIG. 3. (a) The measured peak position of resonance B (blue data points) and different stray field components as a function of the magnetic field direction $\phi$. The experimental data are fitted using a cosinusoidal function (blue curve). Inset: example of the measured current $|I_{SD}|$ (violet circle) and the fitted Gaussian function (red curve) as a function of the scanning magnetic field strength $B_{ext}$, with the field direction at $\phi = 325°$. The nonzero background current of $|I_{SD}|$ in the inset is most likely caused by high microwave power. (b) Plot of both the experimental (blue data points) and simulated (considering different transverse magnetic field gradients) resonance amplitude $I_p$ of resonance B as a function of the magnetic field direction $\phi$. The scaling factor of $C = 1.9$ is used in Eq. (1) for the calculation of all the simulated curves.

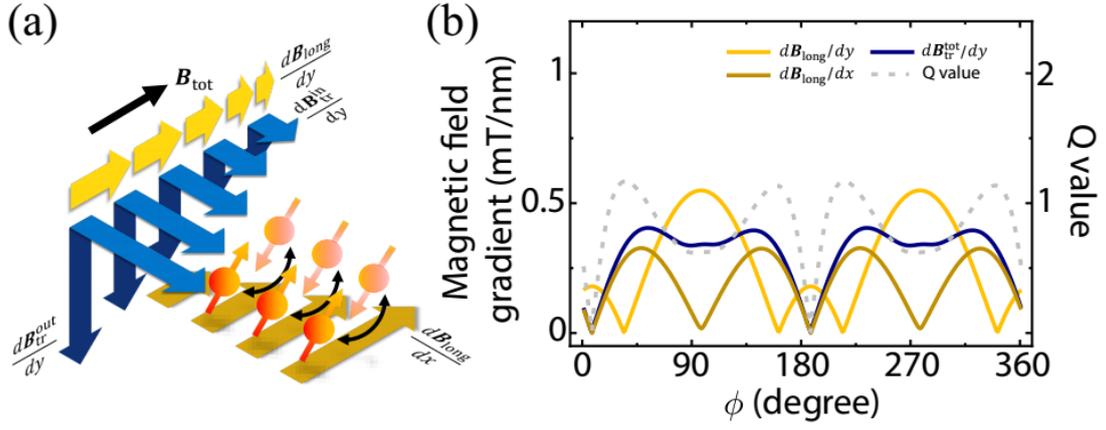

FIG. 4. (a) Illustration of different magnetic field gradients and their effects on the oscillating electron spin. The transverse magnetic field gradients $d\mathbf{B}_{tr}^{in}/dy$ and $d\mathbf{B}_{tr}^{out}/dy$ enable spin flips when the electron is driven by the oscillating microwave fields. The longitudinal field gradients $d\mathbf{B}_{long}/dy$ and $d\mathbf{B}_{long}/dx$ lead to spin dephasing and $d\mathbf{B}_{long}/dx$ also introduces spin addressability in our device. (b) Numerically simulated magnetic field gradients and the calculated quality factor $Q$ as a function of the external magnetic field direction $\phi$.

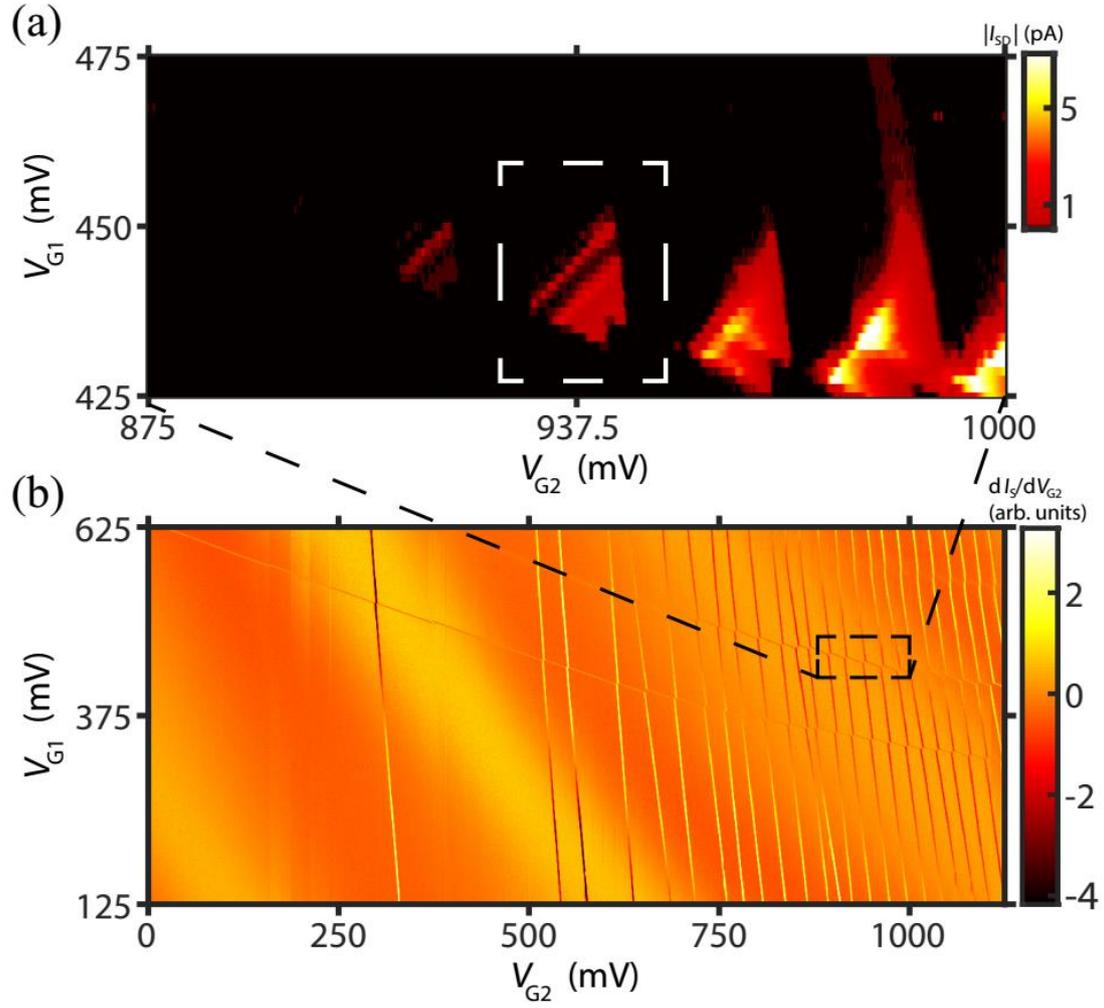

FIG. 5. (a) Transport current $|I_{SD}|$ as a function of $V_{G1}$ and $V_{G2}$ with a bias voltage $V_{SD} = -4$ mV and an external magnetic field $B_{ext} = 1$ T along the $y$-axis. (b) Stability diagram of the measured DQD with charge sensing.

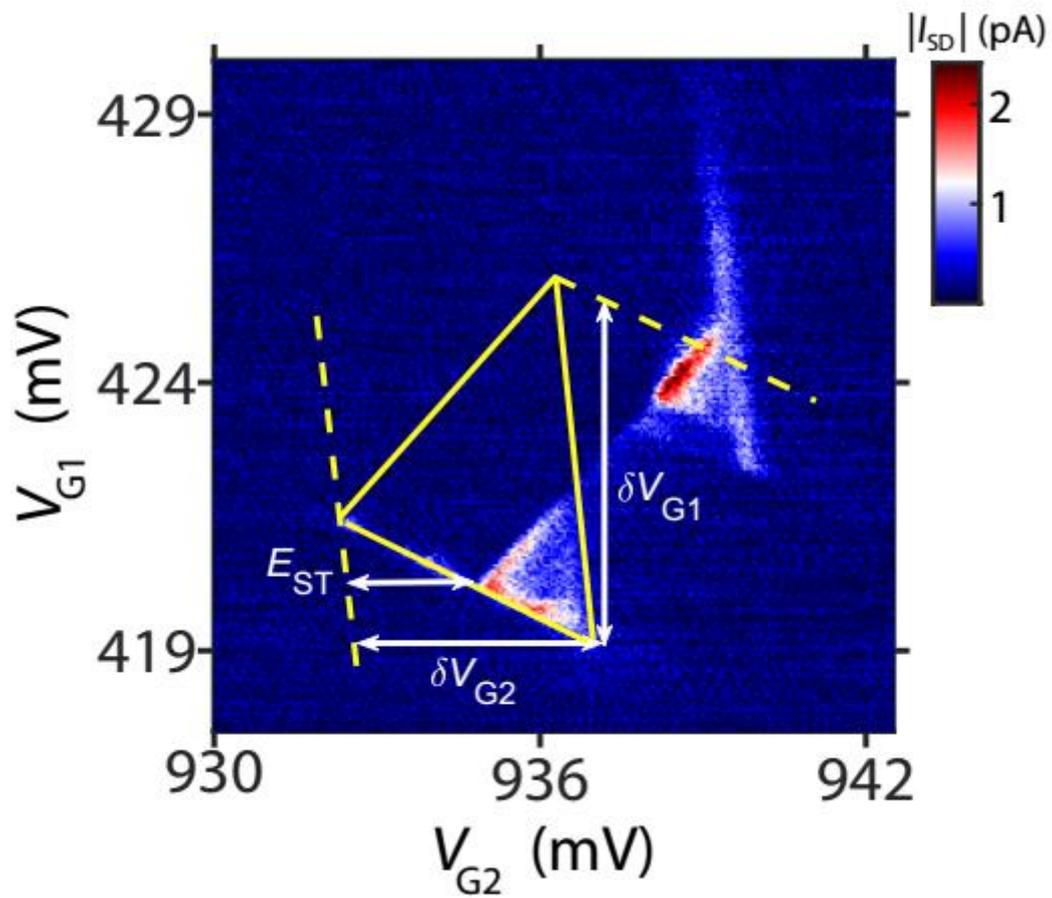

FIG. 6. Illustration of the extraction of the lever arm based on Fig. 1(b) in the main text.

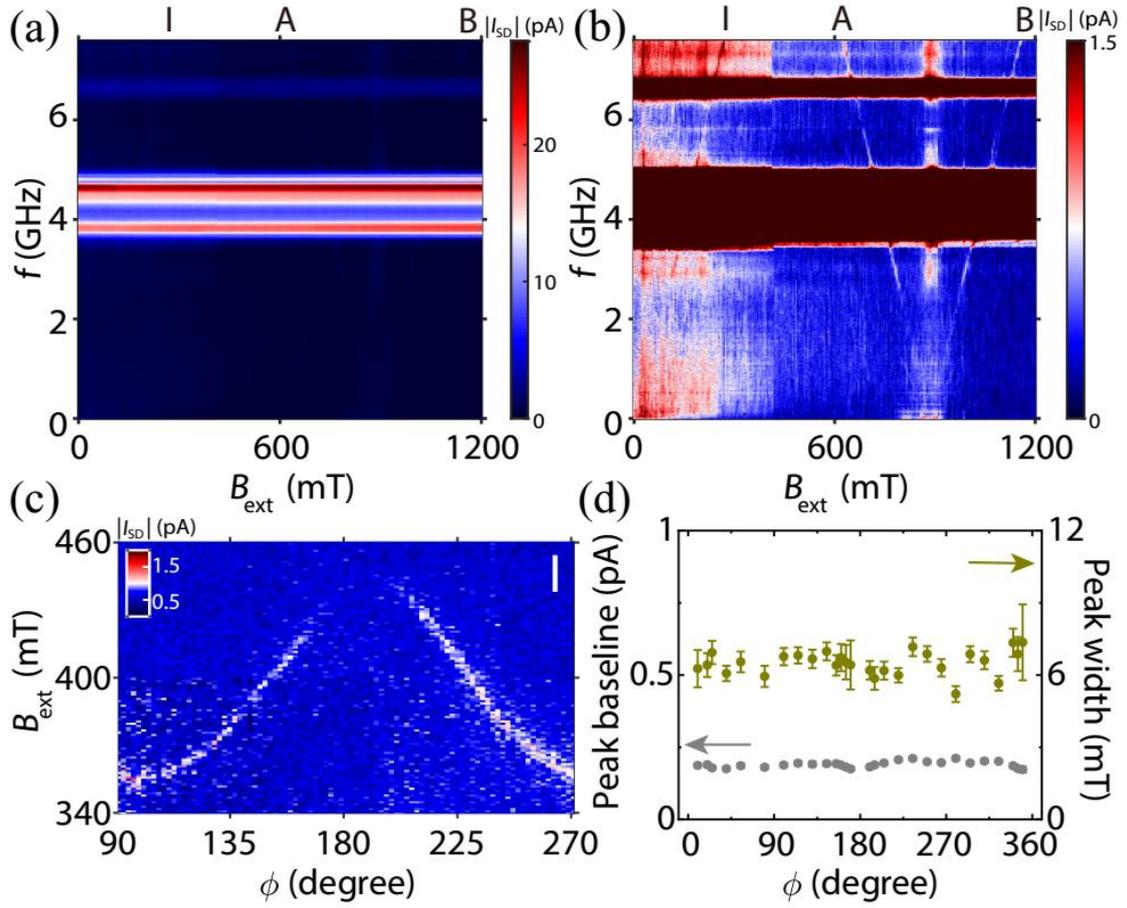

FIG. 7. (a) and (b) show full data of Fig. 2(a), with an additional resonance line I showing the intravalley spin-flip transition. (c) The transport current $|I_{SD}|$ as a function of the magnetic field strength $B_{ext}$ and the magnetic field orientation $\phi$ for the intravelly spin resonance I. Notice the anisotropy magnitude of line I (91 mT) is much smaller than line A (112 mT) and line B (134 mT), which should be attributed to the incomplete magnetization of the micromagnet under lower applied fields. (d) Background leakage current (peak baseline) and measured peak width (full width at half maximum, FWHM) as a function of the in-plane angle $\phi$.

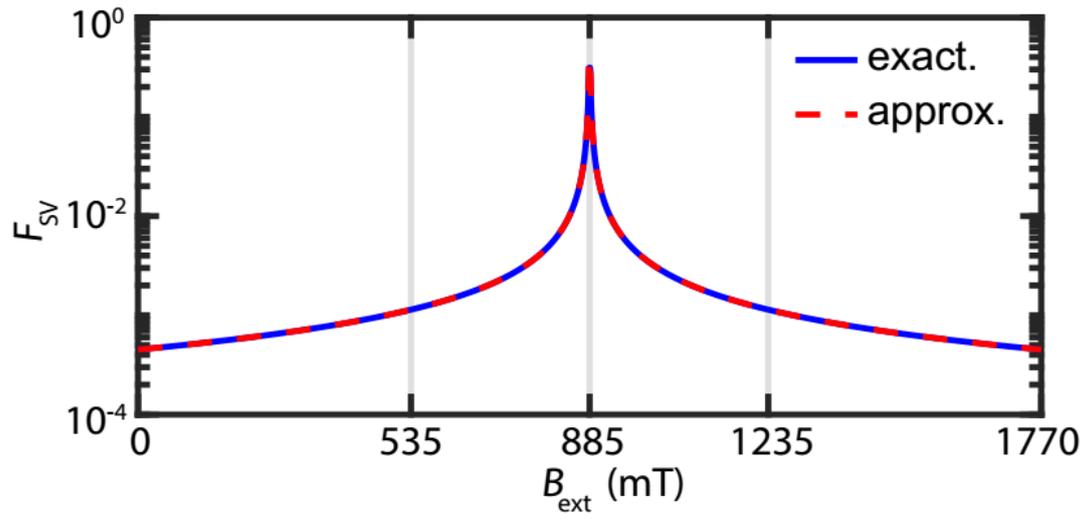

FIG. 8. Comparison of the exact solution (Eq. (12)) and the approximate solution (Eq. (13)) of $F_{SV}$ as a function of the magnetic field strength $B_{ext}$.

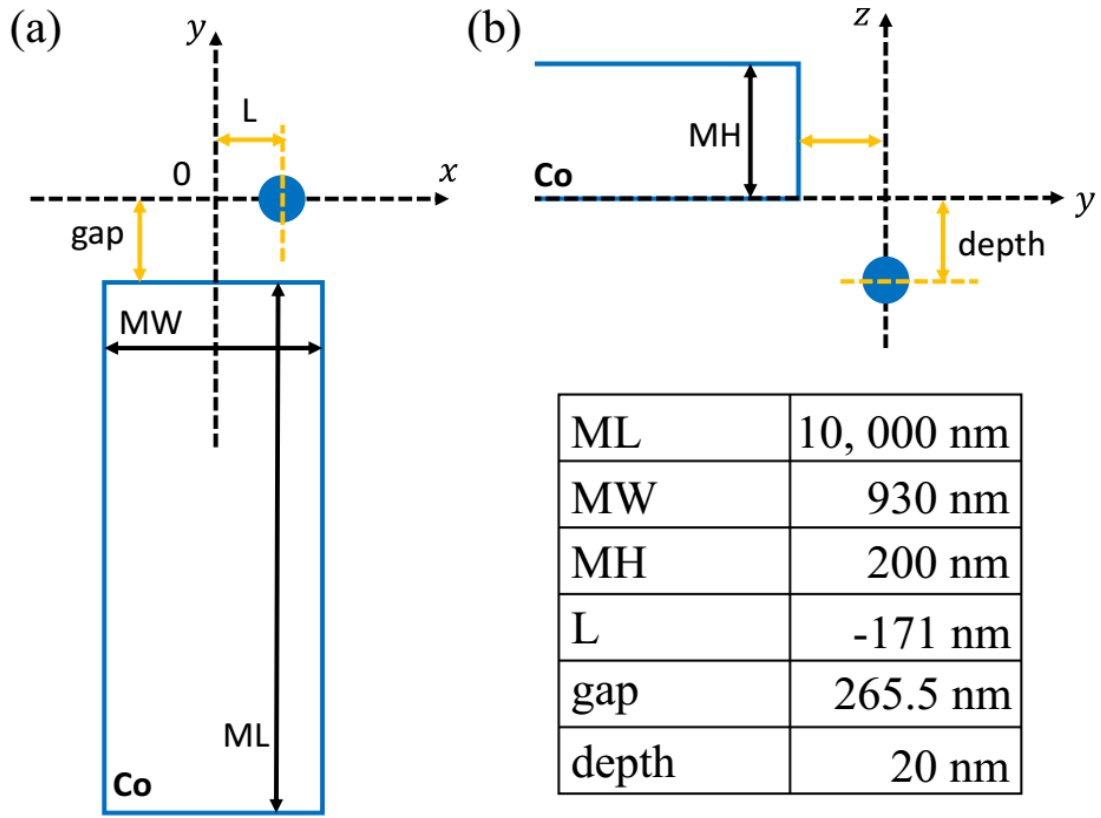

FIG. 9. (a) and (b) are respectively the top view and the side view of the micromagnet with the estimated electron position (blue circle). The Cartesian axis is the same as that in the main text and the size parameters used for simulation are denoted in the table inside the figure.
| ML | 10,000 nm |
| --- | --- |
| MW | 930 nm |
| MH | 200 nm |
| L | -171 nm |
| gap | 265.5 nm |
| depth | 20 nm |

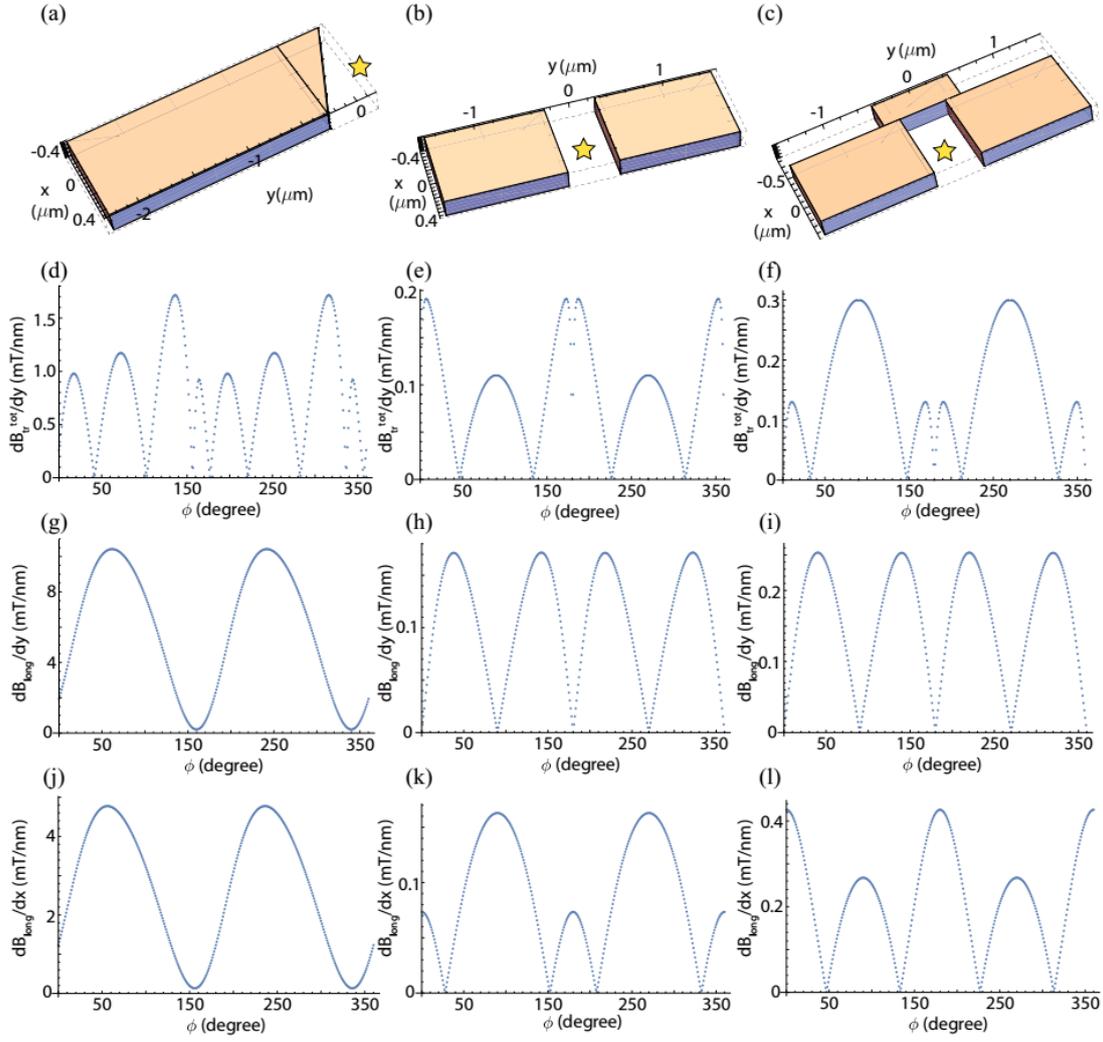

FIG. 10. Diagram of different micromagnet designs (a-c) and the corresponding magnetic field gradients (mT/nm) $d\boldsymbol{B}_{tr}^{tot}/dy$ (d-f), $d\boldsymbol{B}_{long}/dy$ (g-i), $d\boldsymbol{B}_{long}/dx$ (j-l) as a function of the in-plane magnetic field direction $\phi$ (degree). The yellow star shows the position of the electron in the $x$-$y$ plane.